\documentclass[proof]{WileyASNA-v1}

\articletype{Article Type}%

\received{XXX}
\revised{XXX}
\accepted{XXX}

\usepackage{hyperref,natbib,geometry}

\begin{document}

\title{CCD $UBV(RI)_{KC}$ Photometry of the Open Clusters Juchert~9 and Berkeley~97}

\author[1]{{\.I}nci Akkaya Oralhan*}
\author[2]{Ra\'ul Michel}
\author[1]{Yonca Karsl\i}
\author[3]{Hikmet \c{C}akmak}
\author[4]{Hwankyung Sung}
\author[3]{Y\"uksel Karata\c{s}}

\authormark{Akkaya Oralhan \textsc{et al}}

\address[1]{\orgdiv{Department of Astronomy and Space Sciences, Science Faculty}, \orgname{Erciyes University}, \orgaddress{\state{TR-38039, Kayseri}, \country{Turkey}}}

\address[2]{\orgdiv{Observatorio Astron\'omico Naciona}, \orgname{Universidad Nacional Aut\'onoma de M\'exico}, \orgaddress{\state{Apartado Postal 877, C.P. 22800, Ensenada, B.C}, \country{M\'exico}}}

\address[3]{\orgdiv{Department of Astronomy and Space Sciences}, \orgname{Istanbul University Science Faculty}, \orgaddress{\state{34116, \"Universite-Istanbul}, \country{Turkey}}}

\address[4]{\orgdiv{Department of Physics and Astronomy}, \orgname{Sejong University}, \orgaddress{\state{209 Neungdong-ro, Kwangjin-gu, Seoul 05006}, \country{ Korea}}}

\corres{*{\.I}nci Akkaya Oralhan \email{iakkaya@erciyes.edu.tr}}

\abstract{The CCD $UBV(RI)_{KC}$ photometry of the poorly studied open clusters Juchert 9 (Juc~9) and Berkeley 97 (Be~97), which are observed with the 0.84~m telescope at the San Pedro M\'artir National Observatory, M\'exico has been analysed. For the likely cluster members, we determined the reddenings, E(B-V)=0.82$\pm$0.04 (Juc~9) and E(B-V)=0.87$\pm$0.05 (Be~97), from the early type stars. Our distance moduli/distances for only $(B-V)$ colour are (V$_{0}$-M$_{V}$,~d(kpc)) = (13.40$\pm$0.10, ~4.8$\pm$0.2 kpc) (Juc~9) and (12.40$\pm$0.12, 3.0$\pm$0.2 kpc) (Be~97), respectively. The Gaia DR2 distances are d$=$4.5$\pm$1.2 kpc (Juc 9) and  d$=$3.1$\pm$0.7 kpc (Be 97) from the median parallaxes with $\sigma_{\varpi}/\varpi$ $<$ 0.20, which are in good agreement with the photometric distances within the uncertainties. The solar abundance PARSEC isochrones give us the intermediate ages, 30$\pm$10 Myr for Juc~9 and  100$\pm$30 Myr for Be~97.}


\jnlcitation{\cname{%
\author{Akkaya Oralhan~{\.I}},
\author{Michel~R.},
\author{Karsl\i~Y.},
\author{\c{C}akmak~H.},
\author{Sun~H.}, and
\author{Karata\c{s}~Y.}} (\cyear{XXXX}), 
\ctitle{CCD $UBV(RI)_{KC}$ Photometry of the Open Clusters Juc~09 and Be~97}, \cjournal{AN}, \cvol{XXX;XX:X--X}.}

\maketitle

\section{Introduction}
The majority of the stars in the Galaxy are formed in groups as star clusters. One of the groups called open clusters (OCs) can be divided into three age regimes \citep{sun13}; young (Age $<$ 10 Myr), intermediate age (10 Myr $<$ Age $<$ 700 Myr) and old age (Age $>$ 700 Myr). 
Intermediate and old age open clusters play an important role in studying the theories of stellar evolution and chemical evolution of the Galaxy as well as dynamical evolution. OCs have also been used to study the temporal and spatial evolution of the Galaxy \citep{buc14}.
Young OCs give valuable information on the stellar evolution of massive stars, current star formation processes as well as on the evolution of low-mass pre-main sequence (PMS) stars. 
OCs are also powerful probes to study the spiral structure of the Galaxy. It is widely accepted that spiral arms are the preferred sites of star formation. They may be traced by young objects such as giant molecular clouds, H II regions, OB stars, blue and red supergiants and young open/embedded clusters. As massive stars are still in the main sequence or evolved stages, young and intermediate age open clusters are ideal targets of studying the stellar initial mass function in a wide mass range. 

The spiral arm structure of the outer region beyond the Perseus arm is still uncertain. Considering the importance of the spiral arm structure in the outer zones of the Milky Way, we have concentrated poorly studied open clusters, Juchert 9 (Juc 9) and Berkeley 97 (Be 97) as a part of the Sierra San Pedro M\'artir National Astronomical Observatory (SPMO) open cluster survey (\citet{suc07},\citet{tap10}, \citet{akk10},\citet{akk15},\citet{akk19}.)

The two OCs occupy the second Galactic quadrant. Juc 9 lies close to the Outer arm, whereas Be 97 is located at the Perseus spiral arm (Fig.~1).  

We present the astrophysical parameters such as interstellar reddening $E(B-V)$, distance modulus $(V_{0}$--$M_{\rm V})$, distance $d$~(kpc) and Age from our CCD $U\!BV\!(RI)_{KC}$ observations. Gaia DR2 proper motion and parallax \citep{gaia3} are utilise to select the likely members of two OCs.  Note that no spectroscopic observations of the two OCs are available in the literature.
The intermediate age OCs (10 Myr $<$ Age $<$ 700 Myr) (194 OCs) in the second Galactic quadrant is 5$\%$ of 3600 OCs in the catalogue of \cite{kha13}. The distances of 146 OCs are less than 3000 pc, whose distance is similar to Be 97. The rest 48 OCs are at 3--10 kpc. In this sense these new astrophysical parameters of the two OCs  will contribute to the understanding of the structure of second Galactic quadrant.

The equatorial and Galactic coordinates from \cite{kha13} are listed in Table~1. The star charts of Juc~9 (top panel) and Be 97 (bottom  panel) with the field of view of SPM detector (blue rectangle), are displayed in Fig.~2. 

This paper is organized as follows: Section~2 describes the observations and data reductions. The membership selection of the clusters is presented in Section 3. The astrophysical parameters such as  reddening, distance modulus/distance and age are determined in Sections~4 and 5. Discussion and Conclusion are presented  in Section~6. 

\begin{figure}\label{F-figures-01}
	\centering{\includegraphics[width=0.95\columnwidth]{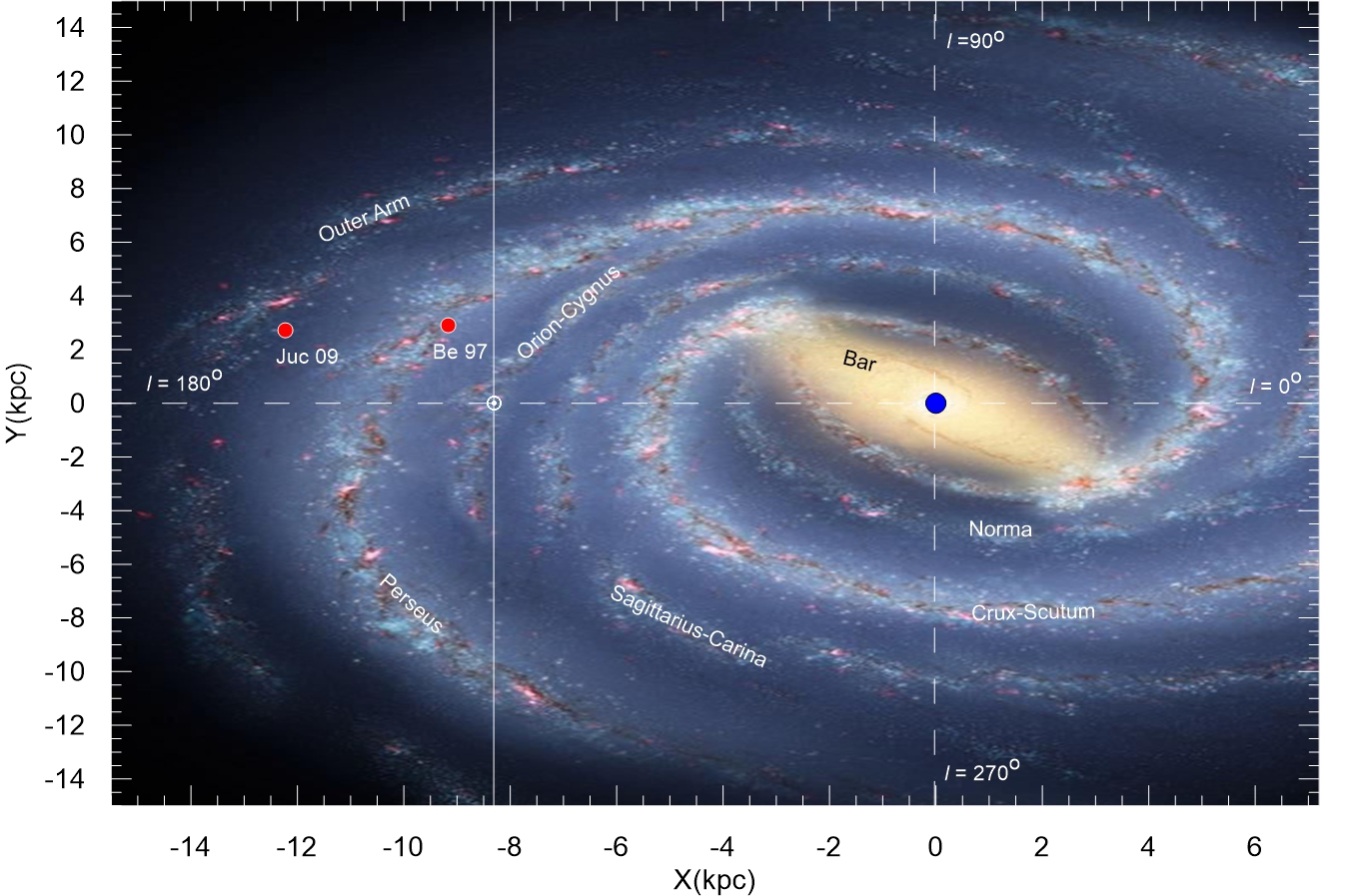}}
	\caption {Spatial distribution $(X,Y)$ (kpc) (filled red dots) of Juc~09 and Be~97. The estimation of $(X,Y)$ (kpc)\protect\footnotemark. The image is adapted from https://www.universetoday.com/102616/our-place-in-the-ga\-lac\-tic-neighborhood-just-got-an-upgrade, credit by Robert Hurt, IPAC; Bill Saxton, NRAO/AUI/NSF.}
\end{figure}
\footnotetext{The $(X, Y)$ (kpc) locations in the Galactic plane are calculated from  their Galactic coordinates (Table 1), the Galactocentric distance of the Sun ($R_{\odot}=8.3\pm0.23$ kpc - \citep{bru11}) and the photometric distances of the two OCs (Table 5). The position of these clusters are plotted in currently accepted spiral arm structure of the Galaxy in Fig.~1.}

\renewcommand{\arraystretch}{1.1}
\begin{table}[!t]\label{Tables-01}
	\caption{Coordinate and observation summary.}
	\setlength{\tabcolsep}{0.22cm}
	\begin{tabular}{lrr}
		\hline
		Cluster  & Juc 09 & Be 97 \\
		\hline
		$\alpha_{2000}$ (h\,m\,s) & 03 55 21 & 22 39 28 \\[1mm]
		$\delta_{2000}$ $(^{\circ}\,^{\prime}\,^{\prime\prime})$ & $+$58 23 30 & $+$58 59 51 \\[1mm]
		$\ell$ $(^{\circ})$, ~b $(^{\circ})$ & 145.12, $+$3.68 & 106.64, $+$0.36 \\[1mm]
		Air mass &1.166 -- 1.200  & 1.297 -- 1.368 \\[1mm]
		Filter U Exp.Time (s) & 80, 900 & 30, 900 \\
		Filter B Exp.Time (s) & 50, 500, 400 & 2,10, 400 \\
		Filter V Exp.Time (s) & 30, 200, 200 & 2, 30, 200 \\
		Filter R Exp.Time (s) & 20,120 & 1,20,100 \\
		Filter I \, Exp.Time (s)&18, 120, 100 & 2x2,20,100 \\
		\hline
	\end{tabular}
\end{table}

\begin{figure}[!h]\label{imagefigures-02}
	\centering{\includegraphics[width=0.6\columnwidth]{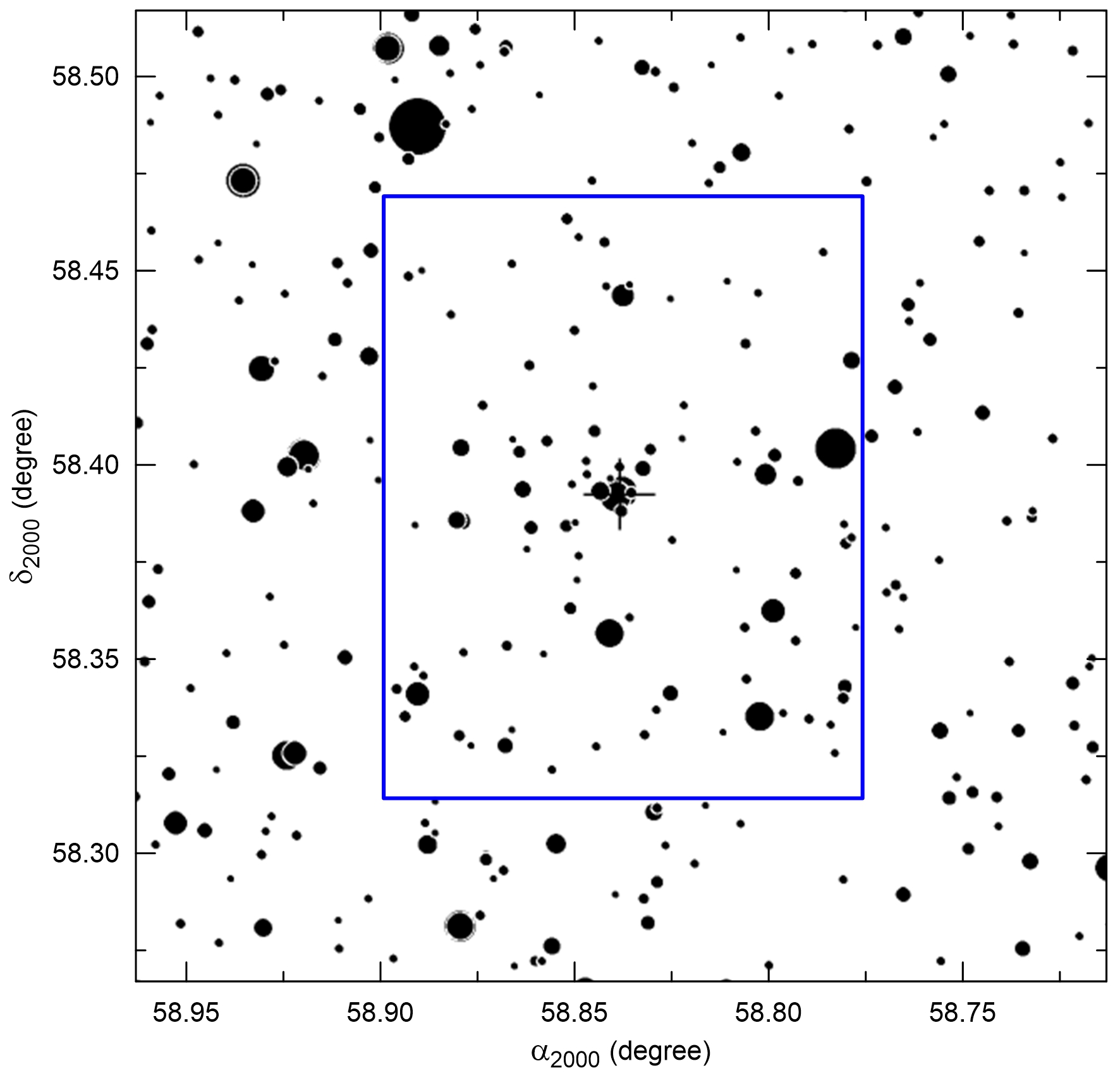}\\[3mm]
		\includegraphics[width=0.6\columnwidth]{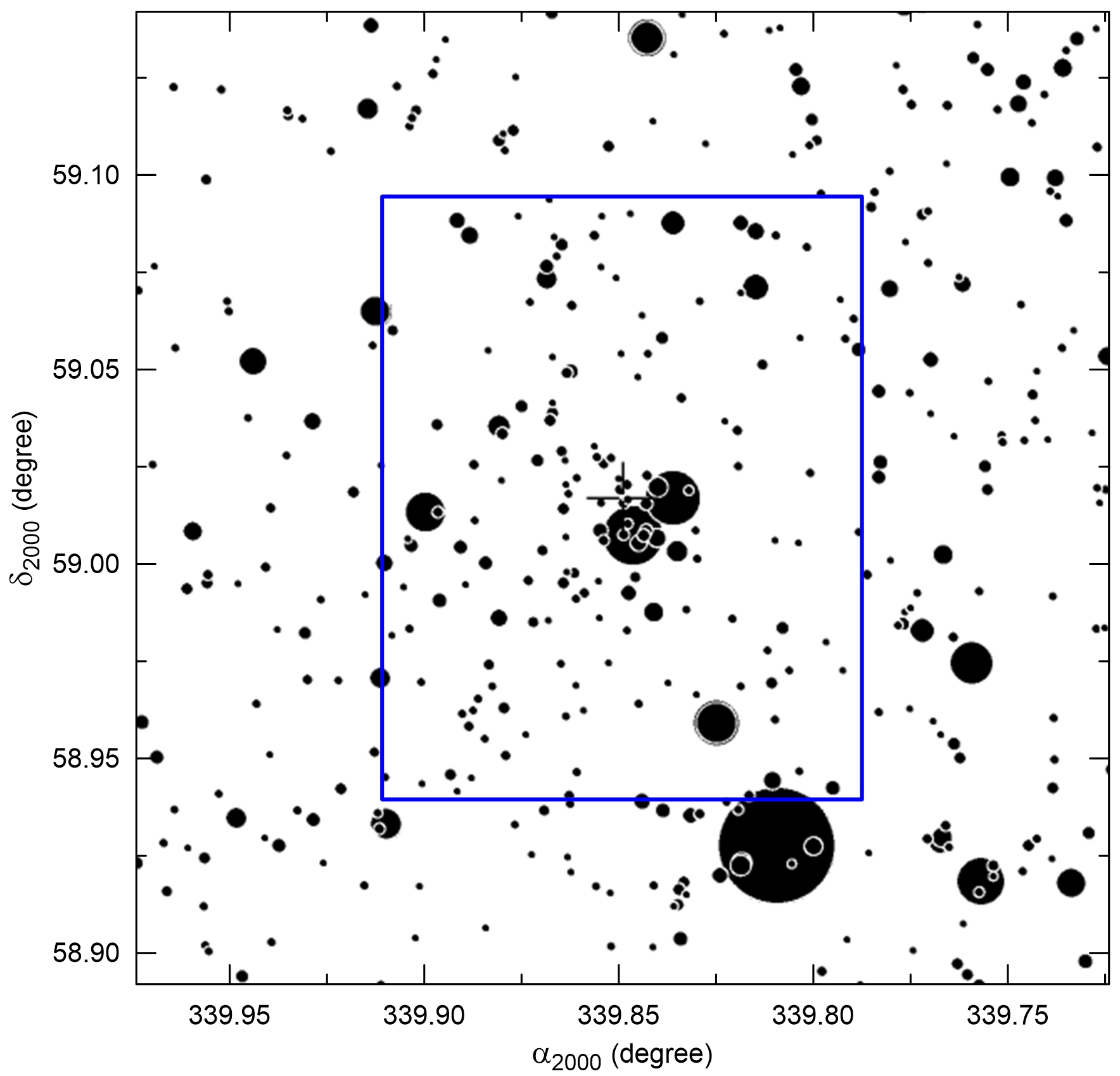}}
	\caption{The star charts of AAVSO\protect\footnotemark of Juc~9 (top) and Be~97 (bottom) with the field of view of SPM detector (blue rectangle), 7.4 (E-W) $\times$9.3 (S-N) arcmin$^2$. Big plus symbol shows the center of the cluster.}
\end{figure}
\footnotetext{https://www.aavso.org/apps/vsp/}

\renewcommand{\arraystretch}{1.3} 
\begin{table}[t!]\label{Tables-02}
	\centering\caption{Coefficients of the transformation equations.}
	\setlength{\tabcolsep}{0.25cm}
	\tiny
		\begin{tabular}{ccccccc}			
			\hline
			Filter&Colour&$\zeta_{\lambda}$ & $k_{\lambda}$&$\eta_{\lambda}$& $rms$ & points \\
			\hline						
			U&U-B & 2.567$\pm$0.009 & 0.452$\pm$0.005 & $-$0.053$\pm$0.005 & 0.025 &    93 \\
			U&U-V & 2.578$\pm$0.010 & 0.459$\pm$0.006 & $-$0.035$\pm$0.003 & 0.029 &   96 \\
			B&U-B & 1.272$\pm$0.005 & 0.223$\pm$0.003 & $-$0.019$\pm$0.003 & 0.014 &   93\\
			B&B-V & 1.287$\pm$0.006 & 0.224$\pm$0.003 & $-$0.03$\pm$0.004  & 0.016 &   119 \\
			V&B-V & 1.330$\pm$0.007 & 0.126$\pm$0.004 & 0.025$\pm$0.004    & 0.019 &   119 \\
			V&V-R & 1.328$\pm$0.007 & 0.127$\pm$0.004 & 0.039$\pm$0.007    & 0.020  &    136 \\
			R&V-R & 1.274$\pm$0.007 & 0.088$\pm$0.004 & $-$0.001$\pm$0.006 & 0.019 &   136 \\
			R&R-I & 1.274$\pm$0.006 & 0.091$\pm$0.003 & $-$0.014$\pm$0.005 & 0.016 &   122\\
			I&V-I & 1.764$\pm$0.008 & 0.073$\pm$0.004 & $-$0.039$\pm$0.004 & 0.021 &   118\\
			I&R-I & 1.754$\pm$0.009 & 0.082$\pm$0.005 & $-$0.090$\pm$0.008  & 0.026 &  122 \\
			\hline	
		\end{tabular}
	\end{table}

\renewcommand{\arraystretch}{1.1}
\begin{table}\label{Tables-03}
	\caption {The mean photometric errors of V, $(R-I)$, $(V-I)$, $(B-V)$ and $(U-B)$ for Juc 9 and Be 97.}
	\setlength{\tabcolsep}{0.20cm}
	{\scriptsize
		\begin{tabular}{cccccc}
			\hline
			&   &    Juc 9  &  \\
			V &<$\sigma_{V}$>&<$\sigma_{R-I}$>&<$\sigma_{V-I}$>&<$\sigma_{B-V}$> &<$\sigma_{U-B}$> \\
			\hline
			12-13 &0.004 &0.005 &0.004 &0.006 &0.005 \\
			13-14 &0.006 &0.005 &0.006 &0.009 &0.008 \\
			14-15 &0.006 &0.003 &0.006 &0.008 &0.006 \\
			15-16 &0.006 &0.003 &0.007 &0.010 &0.008 \\
			16-17 &0.006 &0.003 &0.006 &0.009 &0.011 \\
			17-18 &0.007 &0.004 &0.007 &0.009 &0.022 \\
			18-19 &0.007 &0.005 &0.008 &0.017 &0.049 \\
			19-20 &0.011 &0.008 &0.011 &0.040 &  - \\
			20-21 &0.030 &0.017 &0.035 &  -   &  - \\
			\hline
			&   &    Be 97  &  \\
			\hline
			13-14 &0.001 &0.006 &0.006 &0.003 &0.003 \\
			14-15 &0.003 &0.006 &0.007 &0.010 &0.010 \\
			15-16 &0.003 &0.006 &0.007 &0.005 &0.004 \\
			16-17 &0.004 &0.007 &0.008 &0.004 &0.004 \\
			17-18 &0.005 &0.009 &0.010 &0.008 &0.017 \\
			18-19 &0.024 &0.012 &0.015 &0.026 &0.060 \\
			19-20 &0.014 &0.015 &0.022 &0.071 &  - \\
			20-21 &0.058 &0.020 &0.062 &0.148 &  - \\
			\hline
		\end{tabular}
	} 
\end{table}

\section{Observations and Data Reduction}
Observations of Juc 9 and Be 97 were carried out at the SPMO, during photometric nights on
the date of June 7-10, 2013 UT using the 0.84-m (f/15) Ritchey-Chretien telescope equipped with the
Mexman filter wheel and the ESOPO CCD detector. Seeing was very good ($0^{\prime\prime}$.60 in V long exposure image). The ESOPO detector, a 2048x2048
13.5-$\mu m$ square pixels E2V CCD42-40, has a gain of 1.65 e$^-$/ADU and a readout noise 3.8 e$^-$
at  2$\times$2 binning. The combination of telescope and detector
ensures an unvignetted field of view of 7.4$\times$9.3 arcmin$^2$.

Each OC was observed through the Johnson's $UBV$ and the Kron-Cousins' $RI$ filters with short and long exposure times in order to properly record both bright and faint stars in the region. Standard star fields \citep{lan09} were also  observed at zenith $\approx$ 60 degrees and at meridian to properly determine the atmospheric extinction coefficients. Exposure times used for the observations are given in rows 5--9 of Table~1. Flat fields were taken at the begining and end of each night, and bias images were recorded between cluster observations. 
Data reduction was carried out by one of the authors (R.M.) with the IRAF/DAOPHOT\footnote {IRAF is distributed by the National Optical Observatories, operated by the Association of Universities for Research in Astronomy, Inc., under cooperative agreement with the National Science Foundation.} package \citep{stet87}. 

Standard magnitude for a given filter $\lambda$ is obtained using the following relation.

\begin{flalign}
M_{\lambda} &=  m_{\lambda}-k_{\lambda}X + \eta_{\lambda} C +\zeta_{\lambda}
\end{flalign}

\begin{figure}\label{F-figures-03}
	\centering{\includegraphics[width=0.7\columnwidth]{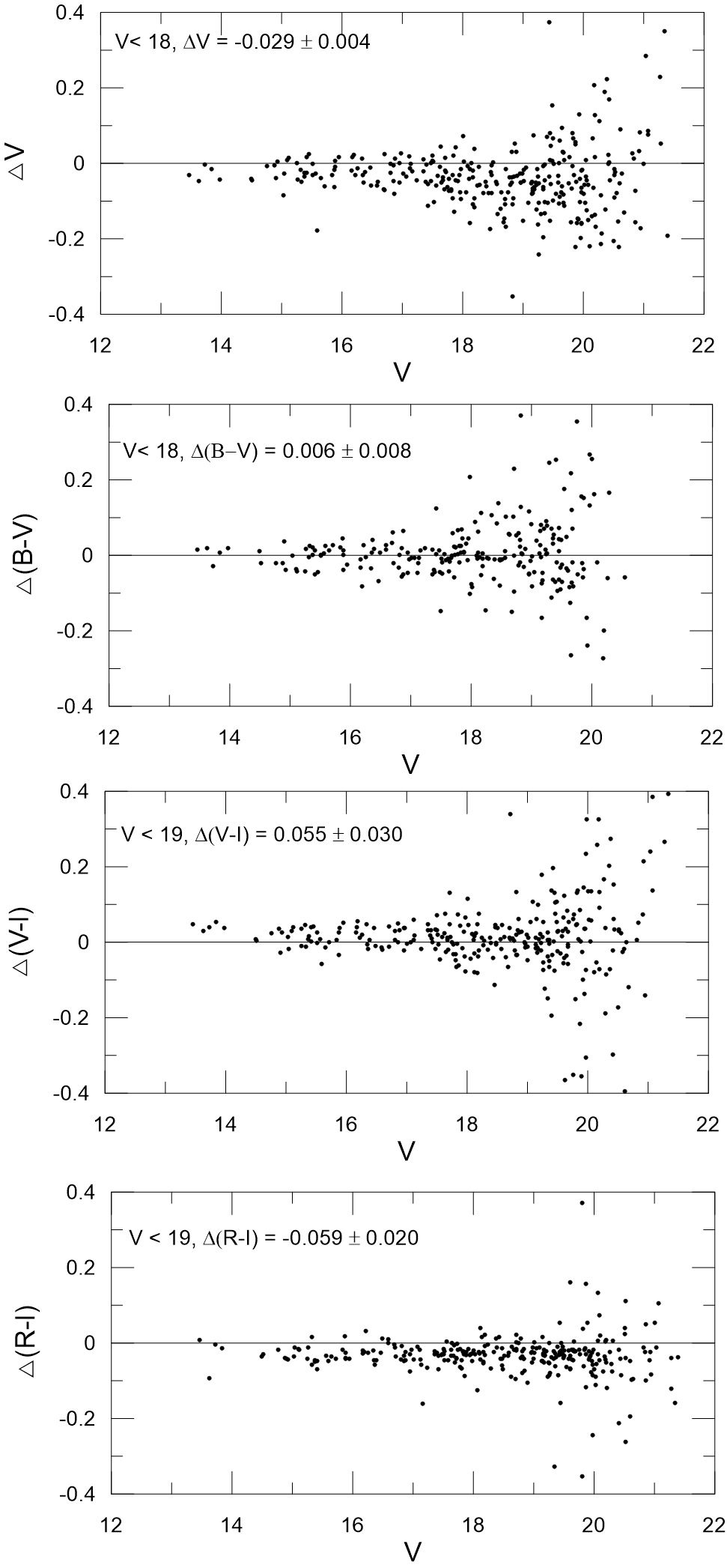}}
	\caption {The differences of $V$-mag and colour indices $(B-V)$, $(V-I)$, $(R-I)$, $(U-B)$ against $V$-mag for Be~97. $\Delta$ means our - \citet{glu13}.}
\end{figure}

\noindent where m$_{\lambda}$, k$_{\lambda}$, C and X are observed instrumental magnitude, extinction coefficients, colour index and air mass, respectively. M$_{\lambda}$, $\eta_{\lambda}$, $\zeta_{\lambda}$ are standard magnitude, transformation coefficient and photometric zero point, respectively. More details on data reduction can be found in the papers of \cite{akk10}, \cite{akk15} and \cite{akk19}. The resulting coefficients for a given filter (Col.~1) with respect to the relevant colour (Col.~2) are listed in Table 2. The rms deviation (Col.~6) and number of stars used in the fit (Col.~7) are also presented in Table 2.

The mean photometric errors in magnitude and colours of the two OCs are listed in Table 3. The photometric errors for faint stars (V $>$ 18 mag) increase rapidly due to the small size of the telescope used in the observation. Our photometry is shallower than 2MASS \citep{skr06}, IPHAS \citep{dre05} and GAIA \citep{gaia1} by about 1 mag, 3 mag and 2 mag, respectively.

Note that there is no reported photometric study on Juc 9. Our photometry for Be~97 is compared with  \citet{glu13}. For common stars, differences of $\Delta V$, $\Delta (B-V)$, $\Delta (V-I)$ and  $\Delta (R-I)$ are displayed in Fig.~3. Here, the difference $\Delta$ is in the sense that our photometry minus \citet{glu13}. For the interval of  $13 < V < 18$, the V magnitudes of \citet{glu13} seem to be slightly fainter. Our $(V-I)$ values are somewhat redder for the same V-mag range. Whereas our (R-I) colour is slightly bluer than \citet{glu13}. For $13 < V < 18$, our $(B-V)$ values are in good consistent with the ones of \citet{glu13}. For $V < 18$ and  $V < 19$, the mean value and standard deviation of the differences are indicated in the panels of Fig.~3.

\begin{figure*}\label{F-figures-04}
	\centering{\includegraphics[width=0.40\textwidth]{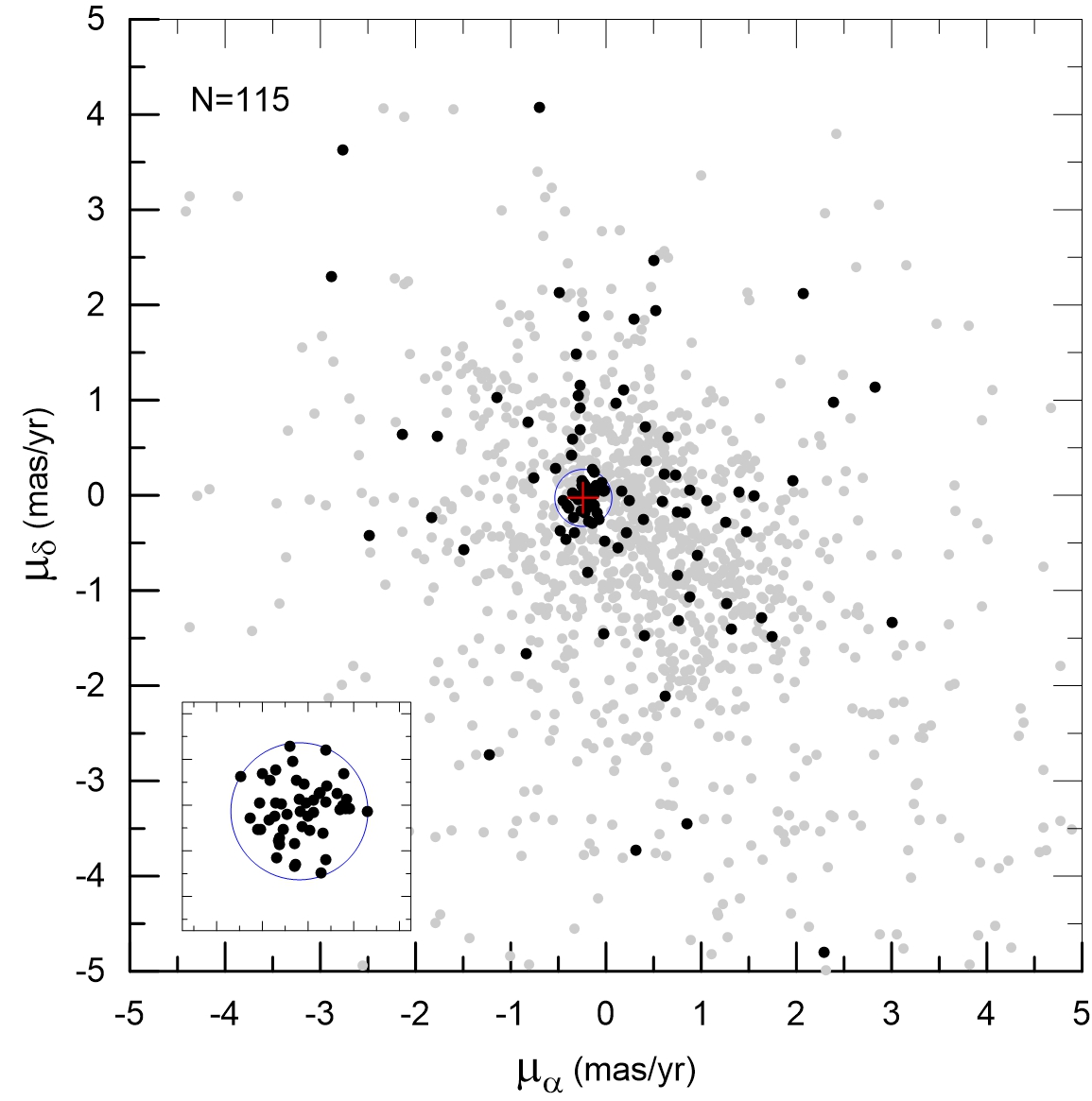} \hspace*{18mm}
		\includegraphics[width=0.40\textwidth]{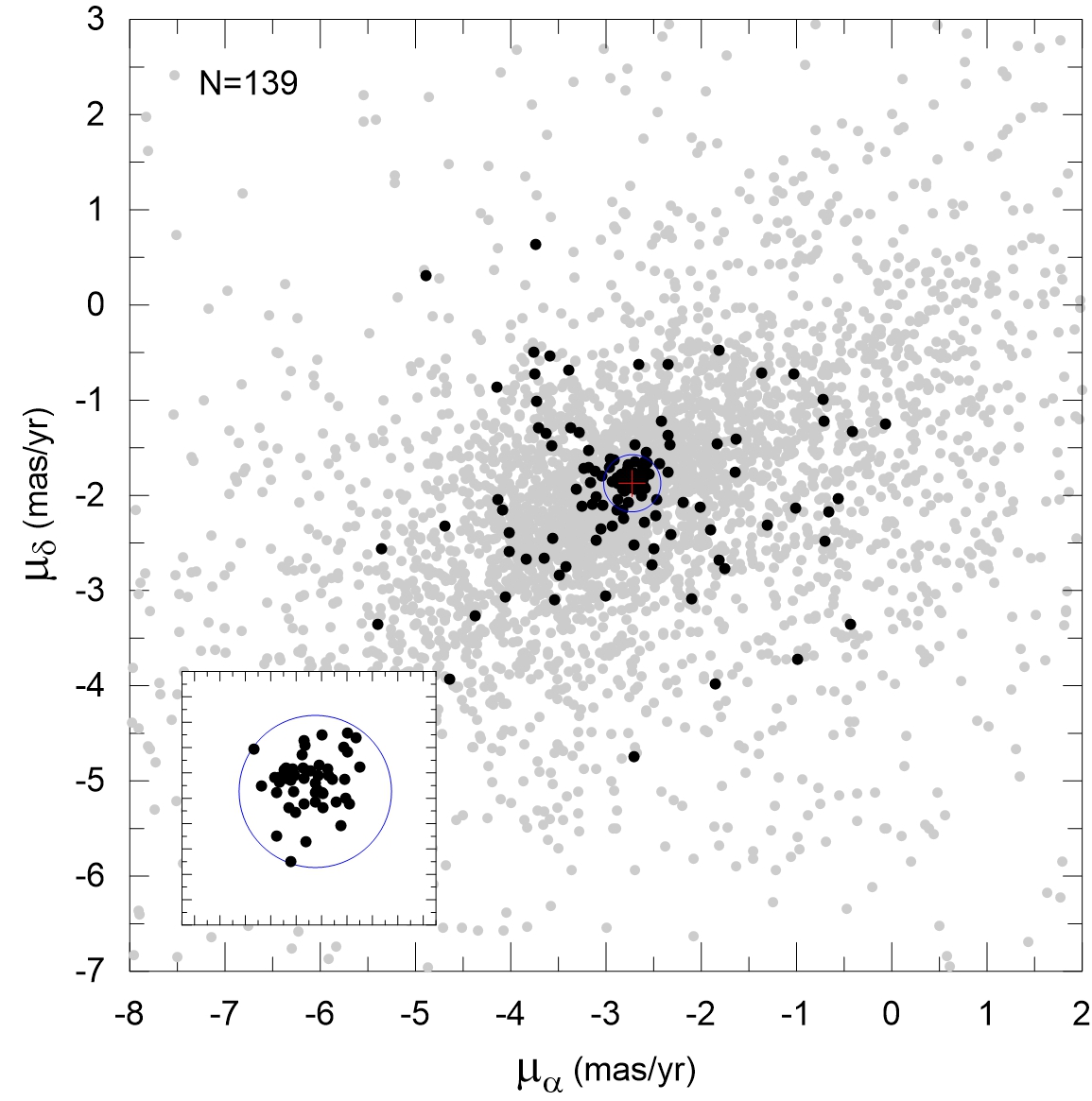}}
	\centering{\includegraphics[width=0.47\textwidth]{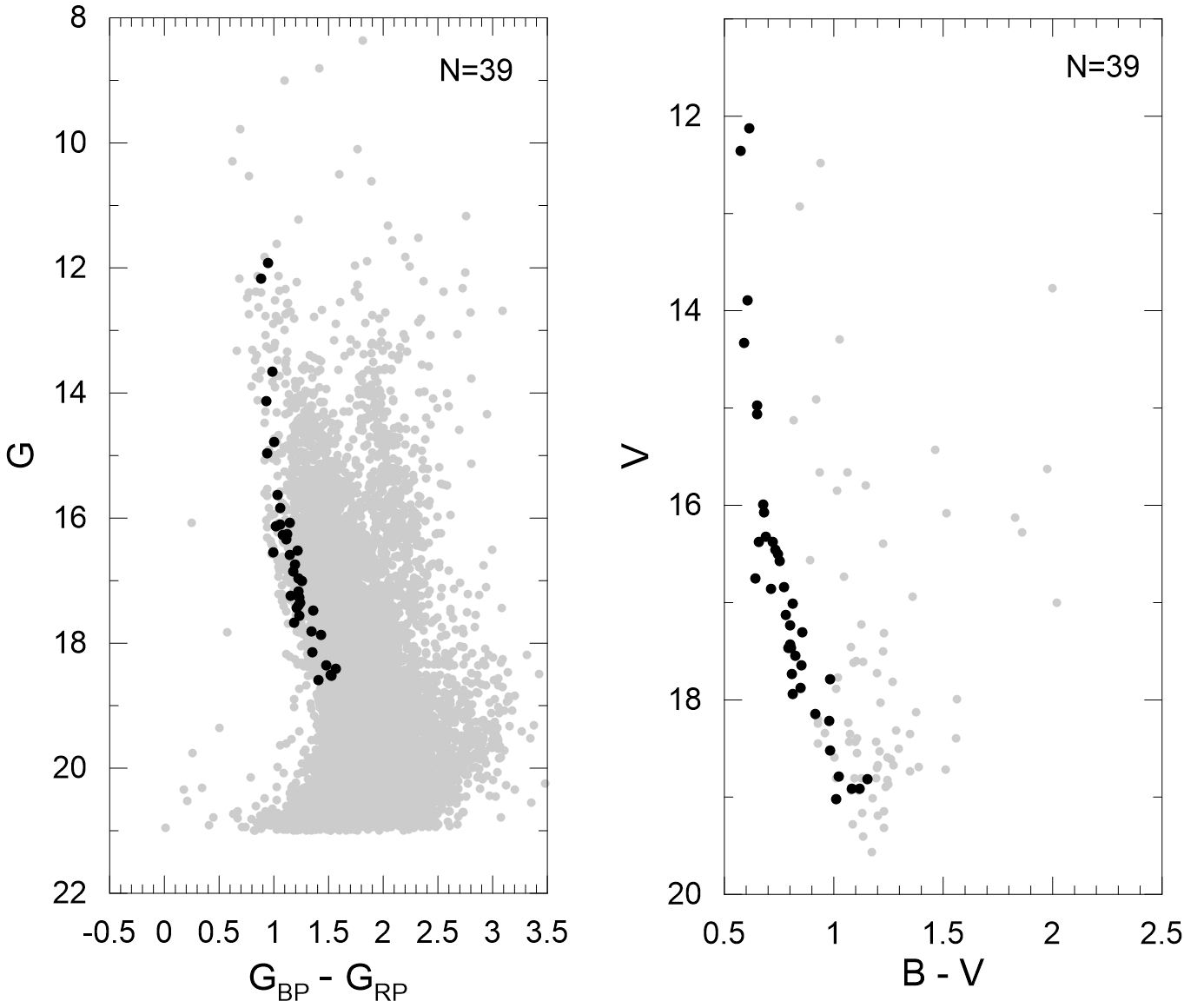} \hspace*{7mm}
		\includegraphics[width=0.47\textwidth]{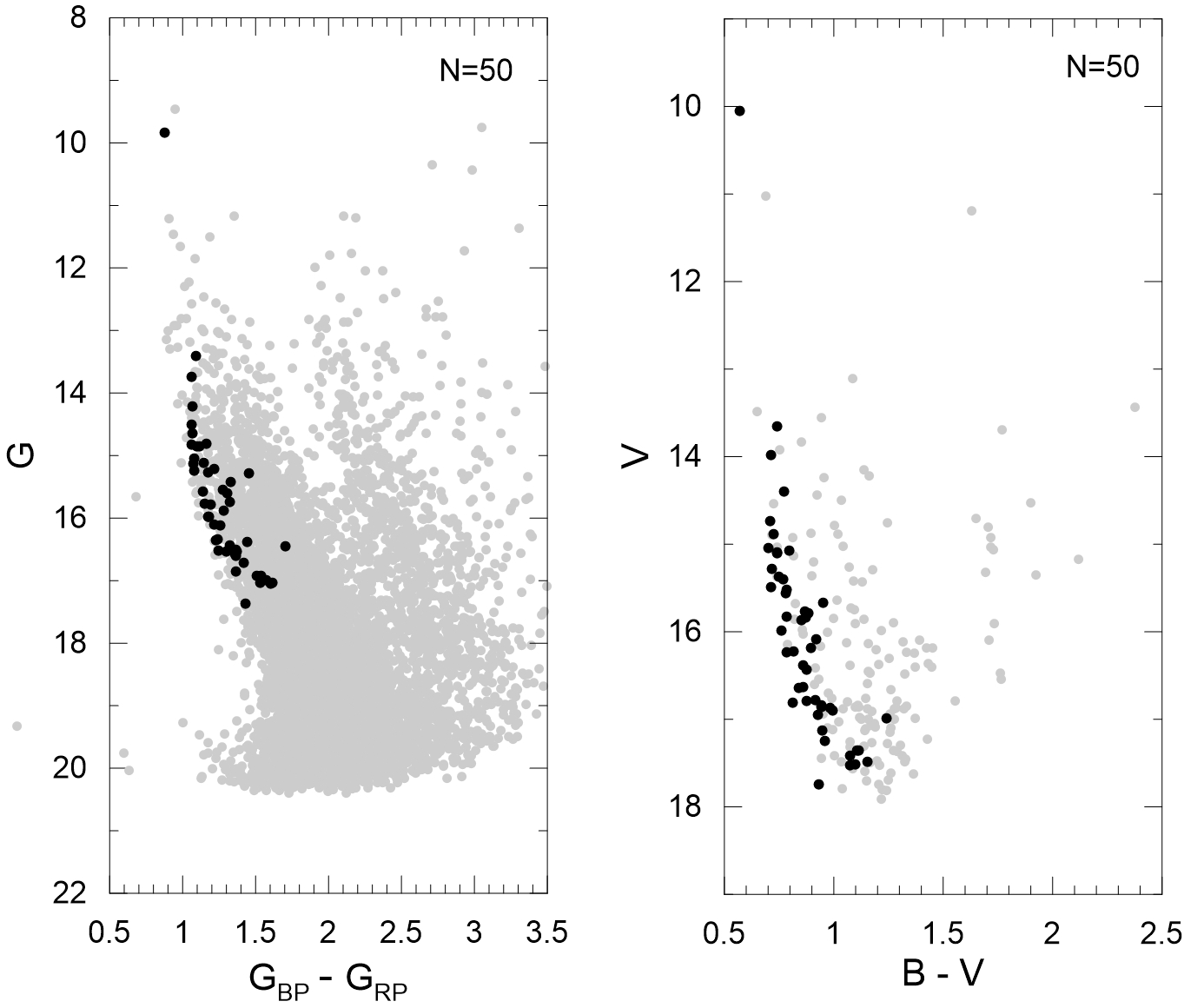}}
	\caption {The $\mu_{\alpha}$ versus $\mu_{\delta}$ for Juc~9 (115 filled dots, left panels) and Be~97 (139 filled dots, right panels). The field stars inside 15 arcmin are shown with small grey dots. The fitted proper motion circles denote the radii of 0.3 mas~yr$^{-1}$, which are considered as the likely members. The big red pluses indicate the median values. A single stellar cluster sequences of the probable members (filled dots of bottom panels) are separated out on $(G, G_{BP}-G_{RP})$ and $(V, B-V)$ diagrams.}
\end{figure*}

\section{Membership selection}
We have utilised Gaia DR2 proper motions \citep{gaia3} to select the probable cluster members of Juc 9 and Be 97. The stars in our photometric catalog are matched with the Gaia DR2 sources.  
The cluster stars of the two OCs are shown as black filled dots on the ($\mu_{\alpha}$,~$\mu_{\delta}$) diagram in Fig.~4.  The members of both OCs are almost clustered around ($\mu_{\rm \alpha},~\mu_{\rm \delta}$)=($-$0.189$\pm$0.094,~$-$0.020$\pm$0.136) and ($\mu_{\rm \alpha},~\mu_{\rm \delta}$)=($-$2.759$\pm$0.102,~$-$1.820$\pm$0.099) mas~yr$^{-1}$, respectively. The fitted proper motion radii of 0.3 mas~yr$^{-1}$ represent well the likely cluster members. These fits have been made by eye on a circle in the $\mu_{\alpha}$ versus~$\mu_{\delta}$ plot until the probable members provide good single stellar sequence on  $(G, G_{BP}-G_{RP})$ and $(V, B-V)$ diagrams (bottom panels of Fig.~4).
The stars within the blue circles are used to calculate both the median  values ($<\mu_{\rm \alpha}>$ and $<\mu_{\rm \delta}>$) and the quantity $\mu_{\rm R}=\sqrt{ (\mu_{\rm\alpha}-<\mu_{\rm \alpha}>)^{2} + (\mu_{\rm \delta}-<\mu_{\rm \delta}>)^2}$). 
The grey dots of Fig.~4 denote the background/foreground field stars for a region (R $=$ 15 arc min) centered on our target OCs. The inset plots show the likely cluster members inside the proper motion circles of Juc~9 (39 members) and Be~97 (50 members). The big red pluses indicate the median values of proper motion components. These likely cluster members are considered for determining the astrophysical parameters on colour-colour and colour-magnitude diagrams of the two OCs.

The angular sizes are determined as  $\theta\sim0.08$ deg (0.0014 in rad) for Juc~9 and $\theta\sim0.07$ deg (0.0012 in rad) for Be~97. The diameter of a cluster can be determined from its distance (Table 5) and angular size. The diameters of Juc~9 and Be~97 are 6.7 pc and 3.7 pc, respectively. In addition, the radius of $\mu=0.3$ mas~yr$^{-1}$ can give the maximum tangential velocity of stars in the cluster. These data can be used to check the stability of the cluster. The maximum tangential velocity of stars in Juc~9 and Be~97 is 6.8 km s$^{-1}$ (Juc~9) and 4.3 km s$^{-1}$ (Be~97), respectively, from the relation, $V_{tan}=4.74\mu\times d(kpc)$. These values indicate that the virial mass of each cluster is about $M_{vir}=36000M_{\sun}$ and $M_{vir}=7900M_{\sun}$, respectively, which are far larger than the cluster masses, 420 $M_{\sun}$ (Juc~9) and  620$ M_{\sun}$ (Be~97), respectively.  The cluster masses are obtained from the PARSEC isochrones \citep{bre12} and Salpeter IMF \citep{sal55}. In this sense they may be in the state of dynamically unstable against the perturbation from the outside. And therefore, the cluster stars may be currently evaporating.

\renewcommand{\arraystretch}{1.1}
\begin{table}[!t]\label{Tables-04}
	\begin{center} 
		\caption {E(V-$\lambda$)/E(B-V) ratios (Col.~2) in terms of four colour indices (Col.~1). R$_{V}$ values are the weighted averages of four colours. Here $\lambda$ is I, J, H and K$_{s}$. N (last column): cluster star numbers.}
		\setlength{\tabcolsep}{0.26cm}
		\begin{tabular}{lllll}
			\hline
			Colour & E(V-$\lambda$)/E(B-V)  & N & E(V-$\lambda$)/E(B-V) & N\\
			\hline
			& Juc 9       &            & Be 97       & \\
			\hline
			V-I      &1.309$\pm$0.058 &13 &1.274$\pm$0.030 &13 \\
			V-J      &2.306$\pm$0.111 &14 &2.243$\pm$0.068 &14 \\
			V-H      &2.607$\pm$0.129 &14 &2.589$\pm$0.083 &14 \\
			V-K$_{s}$&2.799$\pm$0.119 &14 &2.724$\pm$0.107 &14\\[2mm]
			&   R$_{V}$=3.10$\pm$0.07 & 	& R$_{V}$=3.05$\pm$0.05	&    \\
			\hline
		\end{tabular} 
	\end{center} 
\end{table}

\begin{figure}[!t]\label{Fig-5}
	\centering{\includegraphics[width=0.737\columnwidth]{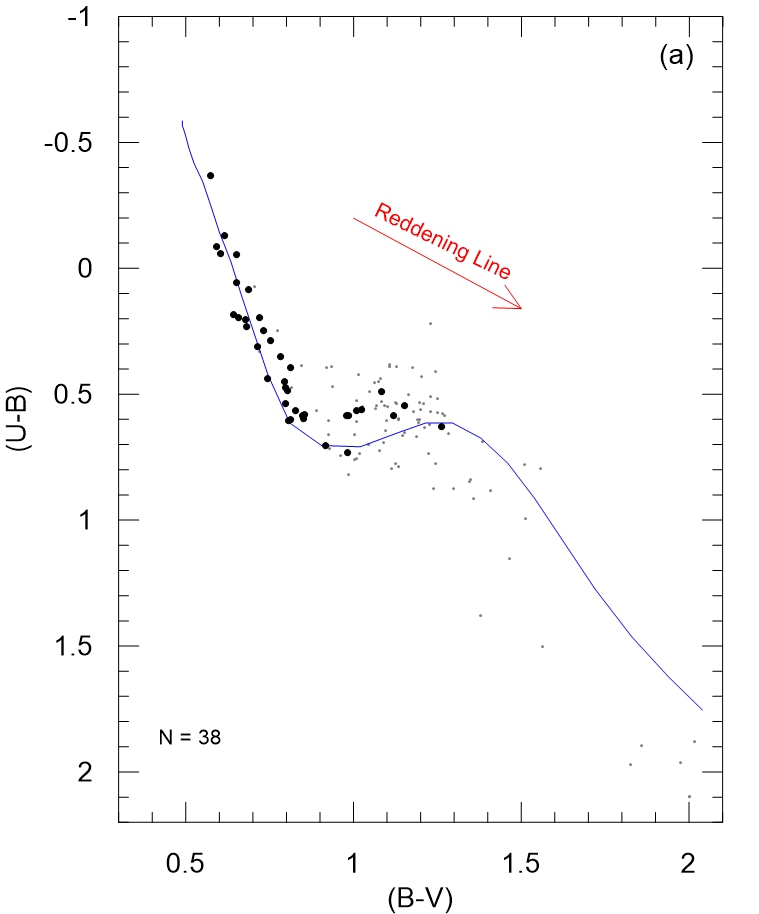} 
		\includegraphics[width=0.695\columnwidth]{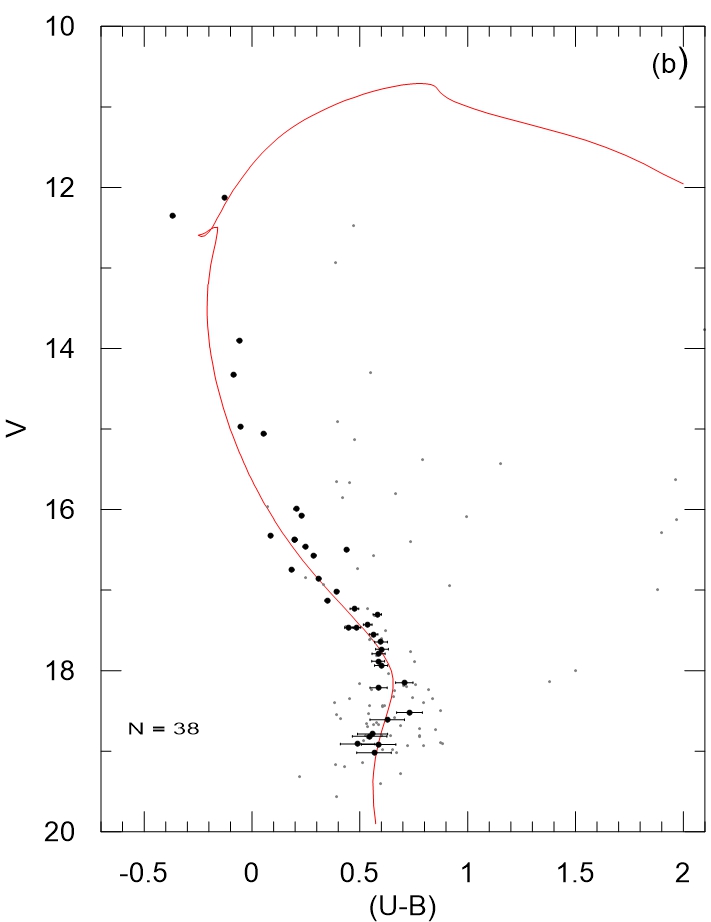}}
	\caption{$(U-B)$, $(B-V)$ (panel a) and $(V,U-B)$ (panel b) of Juc 9. The blue curve shows the reddened  relation of MS stars from SK 82. Filled and grey dots denote the members and non-members, respectively. The arrow denotes the reddening vector. In panel (b) the red curve indicates the reddened solar abundance PARSEC isochrone.}
\end{figure}

\section{Reddenings of Juc~9 and Be~97}
The $(U-B), (B-V)$ (CC) and $(V, U-B)$ diagrams  of Juc~9 and Be~97 are presented in Figs.~5 and 6. It appears that the two OCs contain some early-type stars, which are considered to be members of young open clusters.  From early type stars with $(U-B)<0.30$ on Figs.~5(a) and 6(a), the mean interstellar reddenings are estimated as $E(B-V)$ = 0.82$\pm$0.04 for Juc~9 and  $E(B-V)$=0.87$\pm$0.05 for Be~97, respectively. For these estimates, the intrinsic values of $(U-B)_{0}$, $(V-I)_{0}$, $(R-I)_{0}$, $(V-J)_{0}$, $(V-H)_{0}$, $(V-K_{S})_{0}$ relative to $(B-V)_{0}$ have been used in tables~2--3 of \cite{sun13}. Here we also adopt colour excess ratio $E(U-B)=0.72E(B-V)+0.025E(B-V)^2$ of \cite{sun13}. According to these mean reddening values, the reddened colour sequence of the Schmidt-Kaler (SK82) (blue curves) is fitted to the diagrams of two OCs. According to \cite{gue89}, colour excess ratio of optical-near infrared colours is related to the total-to-selective extinction ratio. We estimated the reddening $E(V-I)$, $E(V-J)$, $E(V-H)$, and $E(V-K_{S}$ using the intrinsic colour relation of O- and B-type stars \citep{sun13}, and then calculated the ratios. The ratios and the derived $R_{V}$ are shown in Table 4. $R_{V}$ is the weighted average of four colour indices. These reddening values have been utilised for deriving the distance moduli (distances) and ages of the two OCs. For $(G_{BP}-G_{RP})$,  we have used the relation of $E(B-V)=0.775E(G_{BP}-G_{RP}$ \citep{bra18}.

Normally $(R-I)$ colour is useful for very red stars with no $(V-I)$ colour. However, the colour excess ratio $E(R-I)/E(B-V)$ is uncertain and its dependence to the reddening law is still unclear. In addition, $R$ magnitude may be affected by $H_{\alpha}$ emission as well. These are the reasons why we do not include $E(R-I)/E(B-V)$ in Table 4. Nevertheless,  we estimate the ratio $E(R-I)/E(B-V)$ as 0.67$\pm$0.03 for Juc~9 and 0.70$\pm$0.03 for Be~97, respectively in order to present the astrophysical parameters for $(R-I)$ colour.

The solar abundance PARSEC isochrone (red curve) of \cite{bre12} on $(V, U-B)$ (panels b of Figs.~5 and 6) is fitted to the lower ridgeline of main-sequence (MS) band and turn-off point. However, the distribution is affected by the effects of evolution and binarity.  The likely cluster members seem to be quite close to the reddened isochrone. 

\begin{figure}[!t]\label{Fig-6}
	\centering{\includegraphics[width=0.74\columnwidth]{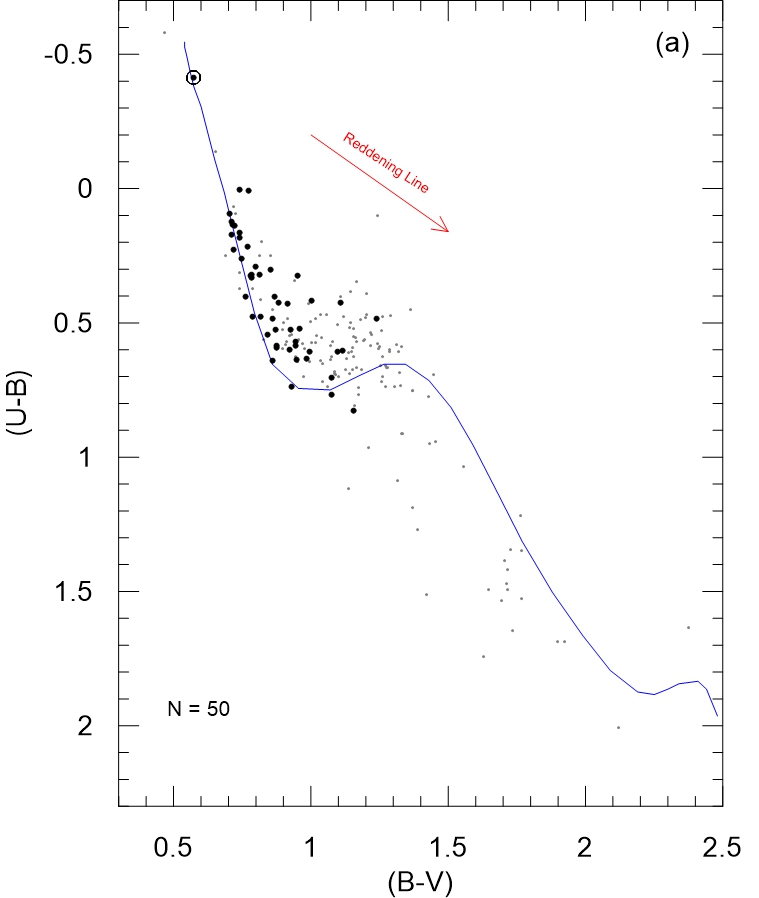} 
		\includegraphics[width=0.70\columnwidth]{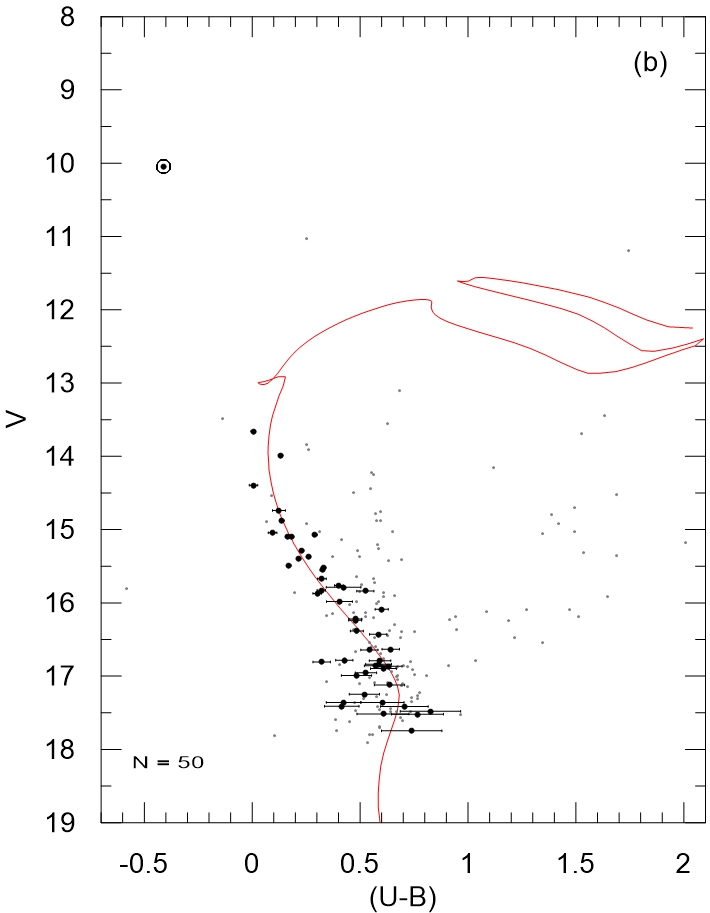}}
	\caption{$(U-B), (B-V)$ (panel a) and $(V, U-B)$ (panel b) of Be 97. The symbols are the same as Fig.~5. The filled circle denotes the star, HD 240015.}
\end{figure}

\begin{figure}[!h]\label{Fig-7}
	\centering{\includegraphics[width=0.72\columnwidth]{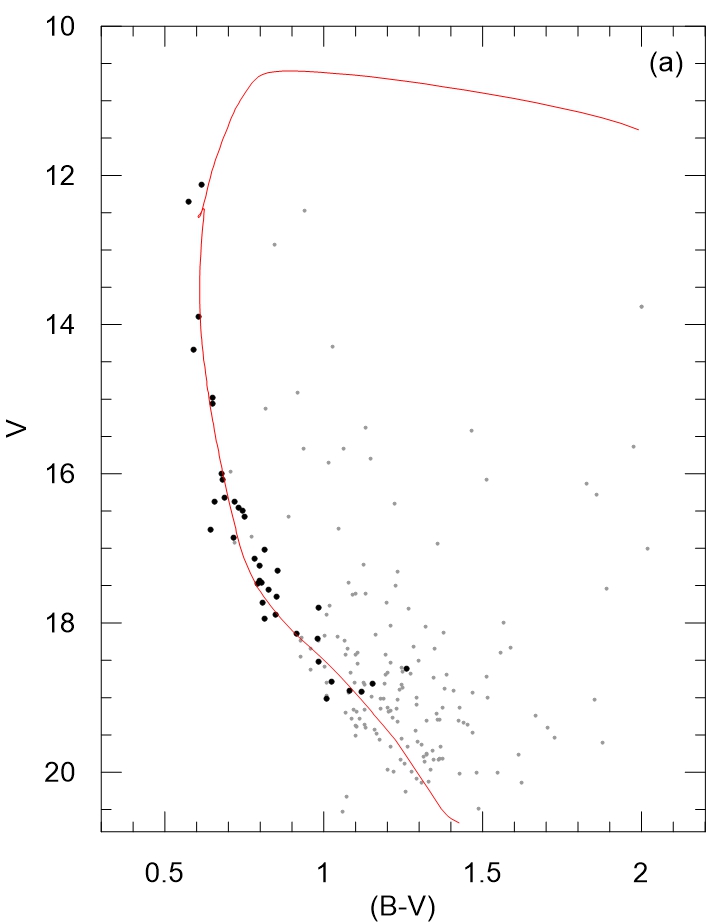}\hspace{3mm}
		\includegraphics[width=0.74\columnwidth]{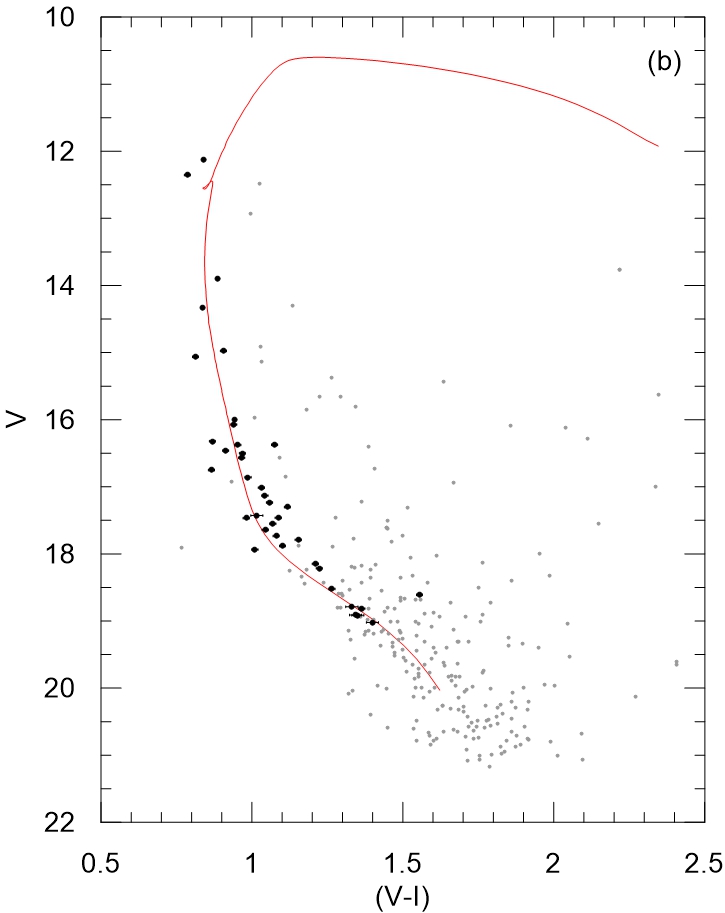}}
	\caption{$(V, B-V)$ (panel a) and $(V, V-I)$ (panel b) diagrams of Juc 9. Solid red curves show the PARSEC isochrones \citep{bre12} with Z =+0.015. Filled and grey dots show the members and the non-members, respectively.}
\end{figure}

\begin{figure}[!t]\label{Fig-8}
	\centering{\includegraphics[width=0.76\columnwidth]{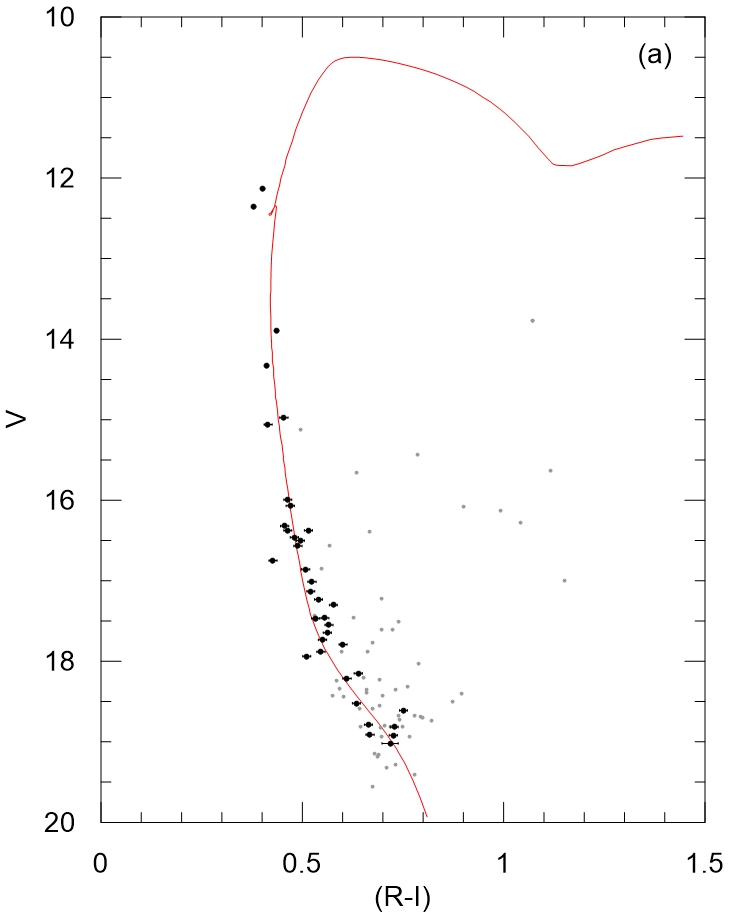} \hspace{5mm}
		\includegraphics[width=0.73\columnwidth]{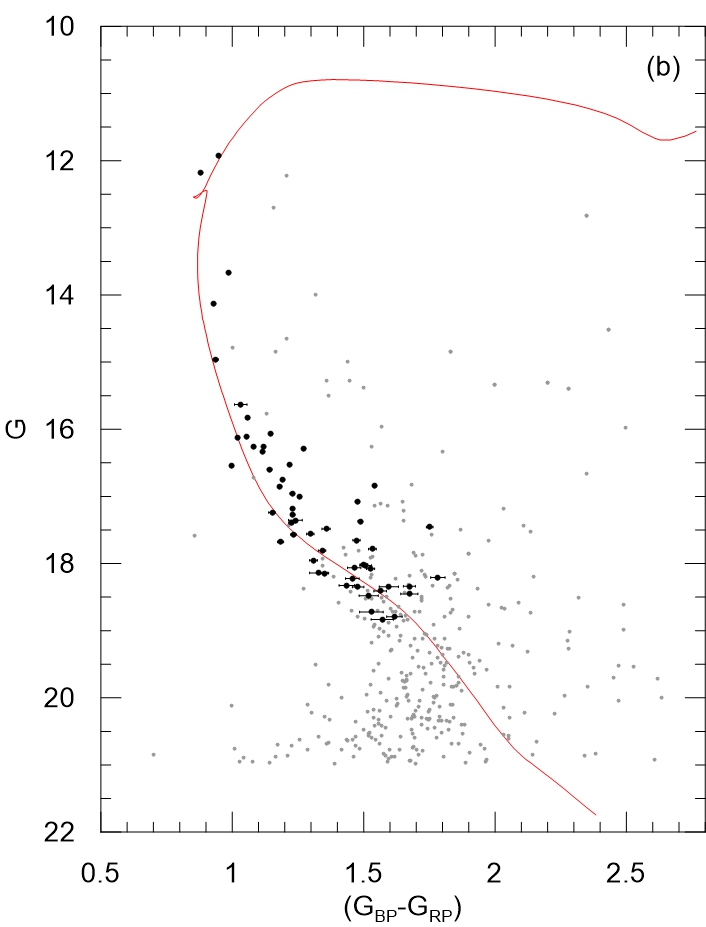}}
	\caption{$(V, R-I)$ (panel a) and $(G, G_{BP}-G_{RP})$ (panel b) diagrams of Juc 9. 
		The symbols are the same as Fig.~7.}
\end{figure}

\begin{figure}[!h]\label{Fig-9}
	\centering{\includegraphics[width=0.74\columnwidth]{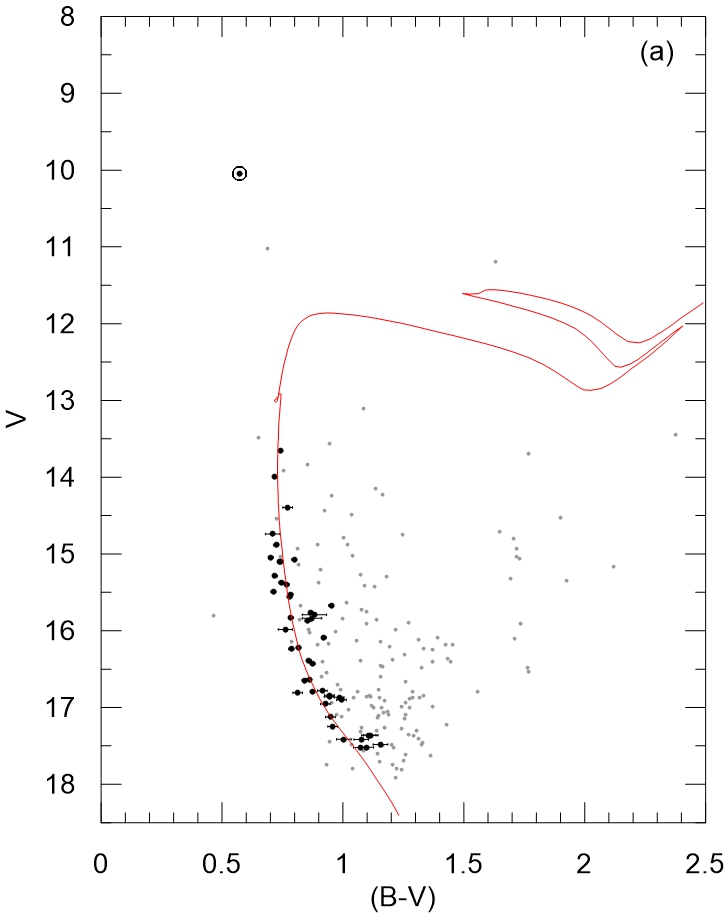} \hspace{5mm}
		\includegraphics[width=0.74\columnwidth]{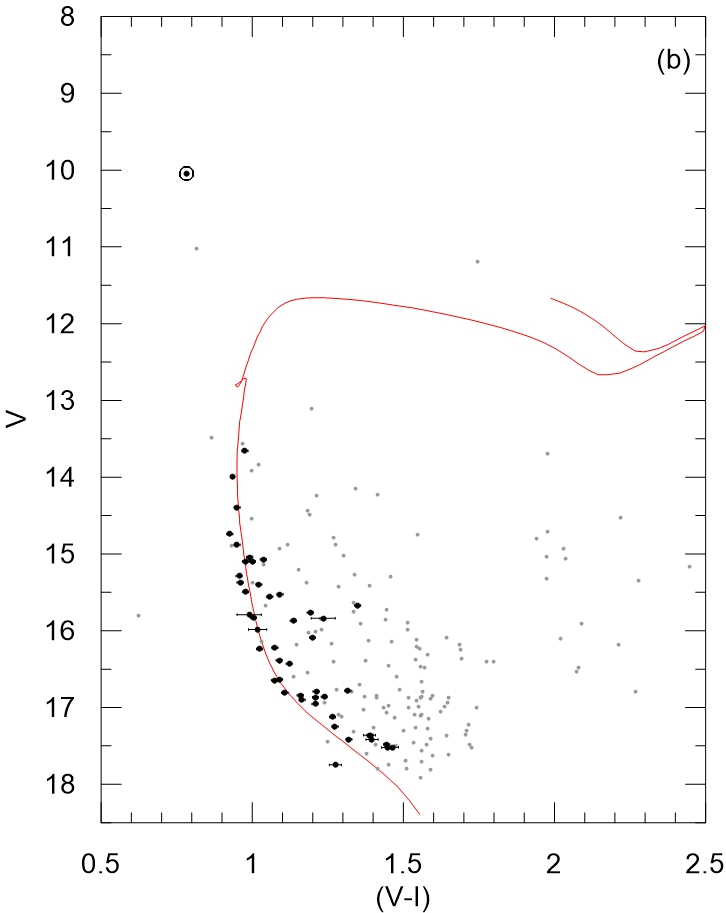}}
	\caption{$(V, B-V)$ (panel a) and $(V, V-I)$ (panel b) diagrams of Be 97. 
		The symbols are the same as Fig.~7. The filled circle denotes the star, HD 240015.} 
\end{figure}

\begin{figure}[!t]\label{Fig-10}
	\centering{\includegraphics[width=0.74\columnwidth]{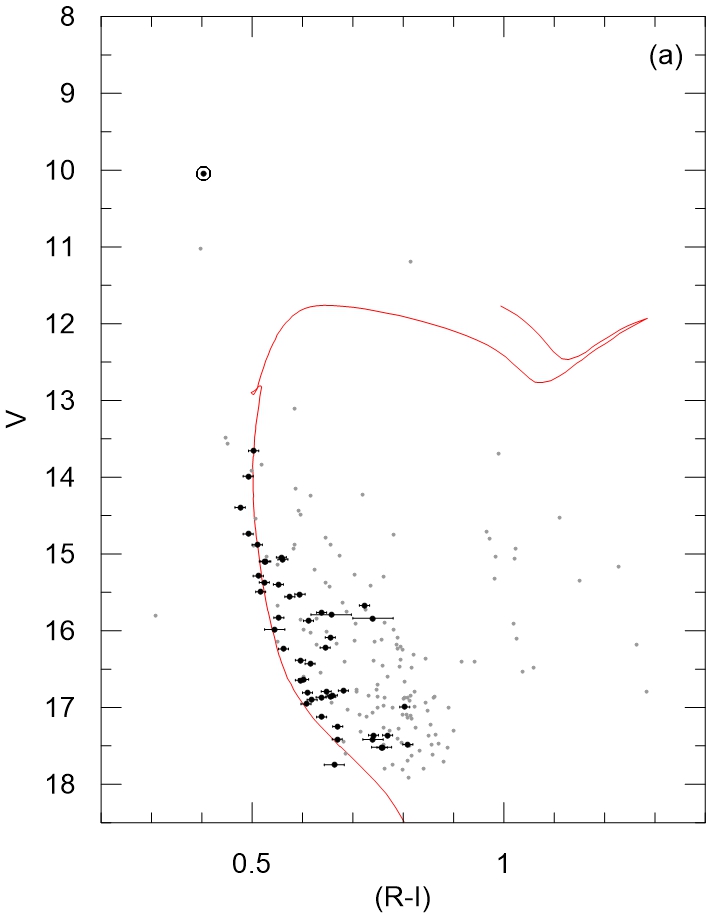} \hspace{5mm}
		\includegraphics[width=0.74\columnwidth]{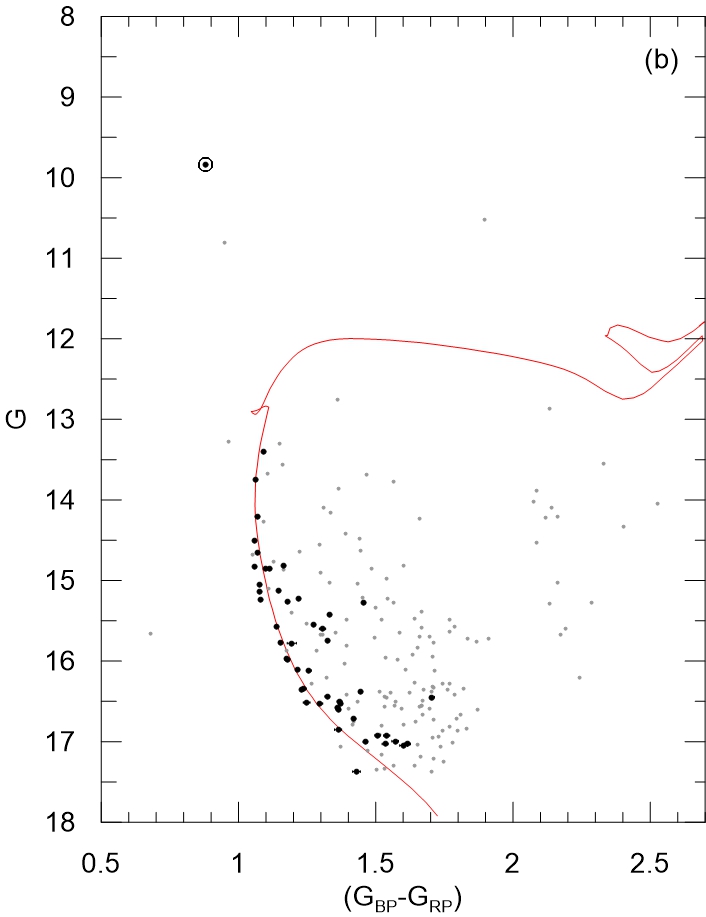}}
	\caption{$(V, R-I)$ (panel a) and $(G, G_{BP}-G_{RP})$ (panel b)  diagrams of Be 97. 
		The symbols are the same as Fig.~7.}
\end{figure}

\renewcommand{\arraystretch}{1.1}
\begin{table}[!h]\label{Tables-05} 
	\centering
	\scriptsize
	\caption{The derived fundamental astrophysical parameters of the two OCs. 
		The reddenings are $E(B-V)$=0.82$\pm$0.04 for Juc 9 and $E(B-V)$=0.87$\pm$0.05 for Be 97. Here we adopt solar abundance value, Z=+0.015.}
	
	\setlength{\tabcolsep}{0.2cm}
	\begin{tabular}{rrrrr}
		\hline
		Juc 9 &$(V_{o}$--$M_{V})$ &  d~(pc) & log(A)  & A~(Myr) \\
		\hline
		(B-V)&13.40$\pm$0.10 &4790$\pm$210 &7.45$\pm$0.20 &30$\pm$10 \\
		(R-I)&13.40$\pm$0.25 &4790$\pm$550 &7.45$\pm$0.30 &30$\pm$10 \\
		(V-I)&13.50$\pm$0.25 &5010$\pm$560 &7.45$\pm$0.30 &30$\pm$10 \\
		(G$_{BP}$--G$_{RP}$)&13.40$\pm$0.30 &4790$\pm$660 &7.60$\pm$0.30 &40$\pm$20 \\
		\hline 
		Be 97& & & & \\
		\hline
		(B-V)&12.40$\pm$0.12 &3020$\pm$170 &8.00$\pm$0.10 &100$\pm$30\\
		(R-I)&12.30$\pm$0.20 &2880$\pm$270 &8.00$\pm$0.10 &100$\pm$30 \\
		(V-I)&12.20$\pm$0.20 &2750$\pm$250 &8.00$\pm$0.10 &100$\pm$30\\
		(G$_{BP}$--G$_{RP}$)&12.30$\pm$0.15&2880$\pm$200&8.15$\pm$0.15&140$\pm$60\\
		\hline
	\end{tabular} 
\end{table}

\section{Distance modulus and age}
To determine the distance moduli (DMs) and ages of Juc 9 and Be 97 from the colour magnitude diagrams (CMDs) (Figs.~7 and 8 and Figs.~9 and 10), the solar abundance (Z $=$+0.015) PARSEC isochrones \citep{bre12} are adopted. The reddened isochrones are calculated according to the reddening and $R_{V}$ obtained above. The PARSEC isochrones are first shifted both vertically and horizontally according to the interstellar reddening. Then the PARSEC isochrones have been shifted vertically to obtain the best fit to the observed main sequence. From this process, we determined the distances of the two OCs. And then we find the best fit isochrone which deliniates the MS turnoff and the location of evolved red giant stars. Different CMDs give slightly different distance moduli. The error in eye-fitting to the MS band is about 0.1 mag. We adopt the error in distance modulus as the quadratic sum of scatter in DMs and eye-fitting error. The error in the age is obtained by jiggling bright/faint-ward the isochrone curve until a good fit of the lower/upper main sequences produce the distance modulus. 
The derived distance moduli, distances (V$_{0}$-M$_{V}$,~d(pc)) and ages are listed  in Table 5. The comparisons of the physical parameters with the literature are given for $(B-V)$ (Table 6). 

We also tried to determine an independent distance using high quality Gaia astrometric data ($\sigma_{\varpi}/\varpi$ $<$ 0.20). The median value of Gaia DR2 parallaxes is 0.223 $\pm$ 0.061 mas (n=5) for Juc 9 and 0.321 $\pm$ 0.067 mas (n=41) for Be 97. Hence the distance from Gaia DR2 to Juc 9 and Be 97 is 4.5 $\pm$ 1.2 kpc and 3.1 $\pm$ 0.6 kpc, respectively.

The Gaia DR2 parallaxes of two brighter stars with V$<$12.50 (Juc 9) are $\varpi$ = 0.223$\pm$0.040 mas (d$=$4.5$\pm$0.8 kpc) and 0.172$\pm$0.04 mas (d$=$5.8$\pm$1.2 kpc), respectively. Their absolute magnitudes are  M$_{V}=-3.95$ and $-$3.60, respectively,  which correspond to a mass $\sim$9.2 M$_{\sun}$ from PARSEC isochrone.

HD 240015 ($V=10.05$, $(B-V)=0.572$, $(U-B)=-0.411$, Sp.~type: B2), the bright blue star in the observed region of Be 97, could be a member of Be 97 according to the star's proper motion. Its membership from its parallax is highly probable considering the error of parallax. 
If the star were a member of Be 97, its age might be about 9 Myr. And the mass of the star from PARSEC isochrone corresponds to 21.4 M$_{\sun}$, i.e. an O-type star. This is inconsistent with the spectral type of the star. In addition, the morphology of MS turn-off between V $=$ 13--15 mag is slightly hooked to the right indicating being an intermediate-age or old open cluster. Therefore, the age of Be~97 from PARSEC isochrone is 100$\pm$30 Myr.	

\renewcommand{\arraystretch}{1.2}
\begin{table*}[!h]\label{Tables-06}
	\centering
	\caption{Comparison with the literature for Juc~9 and Be~97 for only $(B-V)$ colour index.}
	\setlength{\tabcolsep}{0.2cm}
	\resizebox{\textwidth}{!}{
		\begin{tabular}{llllllllll}
			\hline 
			Cluster &E(B-V) &$(V_{0}-M_{V})$&    $d$~(pc) & $Z$   &log Age & Age~(Myr)& Isochrone & Photometry&   Ref.\\
			\hline 
			Juc~9 &0.82$\pm$0.04 &13.40$\pm$0.10 &4790$\pm$210 &+0.015&7.45$\pm$0.20&30$\pm$10&\cite{bre12}  &CCD~$UBV(RI)_{KC}$ &This paper \\
			&0.73  &13.36  &4694$\pm$216 &solar &7.54  &35  &\cite{bon04} &2MASS &\cite{tad12} \\
			&1.42  &12.99 &3971  &solar &6.70  &5  &\cite{gir02}&2MASS     &\cite{kha13}\\
			&0.69  &12.57 &3263$\pm$218&solar&7.75&56&\cite{gir02}& 2MASS&\cite{buk11} \\ 
			&      &      &5376$\pm$2510 &($\varpi$=0.186$\pm$0.087~mas)&  & &   &Gaia DR2 parallax&\cite{can18}\\
			\hline
			Be~97&0.87$\pm$0.05 &12.40$\pm$0.12 &3020$\pm$170 &+0.015&8.00$\pm$0.10&100$\pm$30&\cite{bre12}  &$UBV(RI)_{KC}$ &This paper \\
			&0.77$\pm$0.06 &11.91$\pm$0.19 &2410$\pm$200 &solar&8.40&250&\cite{gir02}  &CCD~$BV(RI)_{C}$ &\cite{glu13} \\
			&0.75  &11.28 &1800$\pm$85 &solar&7.30&20&\cite{bon04} &2MASS &\cite{tad08} \\
			&0.74  &11.28 &1800 &solar&7.55&35&\cite{gir02}&2MASS &\cite{kha13} \\
			&0.73  &12.07 &2589$\pm$174&solar&7.15&14&\cite{gir02}& 2MASS&\cite{buk11}\\ 
			&0.63  &12.24 &2800& &   &               &             &2MASS&\cite{buc13}\\
			&0.59  &11.90 &2400&solar&8.60$\pm$0.10&398$\pm$92&\cite{lej01}&2MASS&\cite{buc14}\\
			&      &      &3185$\pm$680&($\varpi$=0.314$\pm$0.067~mas) &  & &   &Gaia DR2 parallax&\cite{can18}\\
				\hline
		\end{tabular}  
	}
\end{table*}

\section{Discussion and Conclusion}
We present the astrophysical parameters for the poorly studied two OCs in the literature. The reddenings are obtained as $E(B-V)$=0.82$\pm$0.04 for Juc~9 and  $E(B-V)$=0.87$\pm$0.05 for Be~97, respectively. The discrepancies with the literature are at a level of 0.13--0.60 (Juc~9) and 0.10--0.28 (Be~ 97). Note that \cite{kha13} present a large value of 1.42 for Juc 9. However,  our $E(B-V)$ value (Be 97) is very close to the one of \citet{glu13} within the uncertainties. For Be 97, \citet{buc13} and  \citet{buc14} find the values of  $A_{H}=0.34$ from the photometric method and  $A_{H}=0.32$ from the isochrone fitting pipeline, respectively. These total absorptions convert into E(B-V)$=$0.63 and E(B-V)$=$0.59 from the  relation of $A_{H}=0.546E(B-V)$ \citep{dut02}, which is smaller than our values. It is worth to note the reddening determination from optical photometry, especially in the $(U-B, B-V)$ diagram is accurate than that determined from near infrared CMDs.

For four colour indices, $(B-V)$, $(R-I)$, $(V-I)$ and (G$_{BP}$-G$_{RP}$),  distance moduli, distances and ages of the two OCs (Table 5) are in good concordance. For only $(B-V)$, our distance moduli/distances are (V$_{0}$-M$_{V}$,~d(kpc)) = (13.40$\pm$0.10, ~4.8$\pm$0.2 kpc) (Juc~9) and (12.40$\pm$0.12, 3.0$\pm$0.2 kpc) (Be~97), respectively. The differences of distance moduli between ours and literature fall in the intervals of 0.04--0.83 mag (Juc~9) and 0.16--1.12 (Be~97). For the distances, the discrepancies are almost up to level of 0.1--1.5 kpc (Juc~9) and 0.1--1.5 kpc (Be~97). Our distances for Juc 9 and Be 97 are slightly larger than the ones of 2MASS JHK$_{S}$ diagrams and are in good agreements with the GaiaDR2 distances of \cite{can18} within the uncertainties (Table 6).

For $\sigma_{\varpi}/\varpi$ $<$ 0.20, the median Gaia DR2 distances (5 stars) for Juc 9 and  (41 stars) for Be 97 are d$=$4.5$\pm$1.2 kpc ($\varpi$ = 0.223$\pm$0.061 mas) and d$=$3.1$\pm$0.7 kpc ($\varpi$ = 0.321$\pm$0.067 mas), respectively. The Gaia DR2 distances of the two OCs are in good agreement with the photometric ones within the uncertainties (Table 5). 

The Gaia DR2 parallaxes of two brighter stars with V$<$12.50 (Juc 9) are $\varpi$ = 0.223$\pm$0.04 mas (d$=$4.5$\pm$0.8 kpc) and 0.172$\pm$0.04 mas (d$=$5.8$\pm$1.2 kpc), respectively. Their distances guarantee their membership. Their absolute magnitudes are the following, M$_{V}=-3.95$ and  M$_{V}=-3.60$ which correspond to a mass $\sim$9.2 M$_{\sun}$ from PARSEC isochrone.

Although HD 240015 could be a member of Be 97 from Gaia DR2 proper motion and parallax ($\varpi$ = 0.358$\pm$0.031 mas; 2790$\pm$240~pc), we consider that HD 240015 is not a member of Be 97. There are two reasons-(1) there is a big discrepancy in mass from the younger isochrone and its spectral type, and (2) there is a big gap in magnitude/mass between HD 240015 and the next bright star. The MS of Be 97 is curve to the right at V $=$ 13.5--15.0. Therefore, the fitted PARSEC isochrone gives the age of Be 97 as 100$\pm$ 30 Myr.

Age value (30$\pm$10 Myr) of Juc~9 is in reasonable concordance with the ones of the literature (Table 6), except with the 5 Myr value of \citet{kha13}.  Except with the 250 Myr of \cite{glu13} and 398 Myr of \cite{buc14}, our intermediate age value (100$\pm$ 30 Myr) of Be 97 is slightly older than the literature values (Table 6). Discordances between ages of our and the literature are up to 5--26 Myr (Juc~9) and 86--300 Myr (Be~97). Discrepancies of the distance moduli, distances and ages as compared to the literature stem from the usage of different heavy element abundances, isochrones,  reddenings and photometries, as discussed by \cite{moi10}. However, spectroscopic observations are needed for the memberships of brighter stars of the two OCs.

A global systematic offset of Gaia DR2 parallaxes is $\Delta\varpi$ $=$ $-$0.029 mas in terms of an inertial reference frame, emphasized by \cite{gaia5}. Recent values for the zero point shift of parallax are found as  $\Delta\varpi$ $=$$-$0.045$\pm$0.009 mas \citep{yal18}, $\Delta\varpi$ $=$ $-$0.053$\pm$0.003 mas \citep{zin18}, $\Delta\varpi$ $=$ $-$0.046$\pm$0.013 mas \citep{rie18}, respectively.
A correction of 0.005 mas to the median values of our original Gaia DR2 parallaxes give a closer distance with a difference 0.10 kpc (Juc 9) and 0.05 kpc (Be 97).

\section{Acknowledgments}
We thank our referee for the valuable suggestions. We wish to thank the staff of the San Pedro M\'artir Observatory. This work has been supported by Turkish National Science Foundation (TUBITAK), Proje No: 114F123/Program code: 3501. H.S acknowledges the support of the National Research Foundation of Korea (Grant No. NRF-2019RIA2C1009495). We thank William Schuster for personel communication with A. Landolt for the exact defination of filters $(RI)_{KC}$ . This paper has made use of results from the European Space Agency (ESA) space mission Gaia, the data from which were processed by the Gaia Data Processing and Analysis Consortium (DPAC). Funding for the DPAC has been provided by national institutions, in particular the institutions participating in the Gaia Multilateral Agreement. The Gaia mission website is http://www.cosmos.esa.int/gaia.

\end{document}